# Broadband generation of quasi bound-state-in-continuum modes using subwavelength truncated cone resonators

MOHAMAD REZA NURRAHMAN[1], DONGHA KIM[1], KWANG-YONG JEONG[2], KYOUNG-HO KIM[3], CHUN-HO LEE[1,4], MIN-KYO SEO[1,5]

[1] [1]Department of Physics, Korea Advanced Institute of Science and Technology (KAIST), Daejeon, 34141, Rep. of Korea
[2]Department of Physics, Jeju National University (JNU), Jeju, 63243, Rep. of Korea
[3]Department of Physics, Chungbuk National University (CBNU), Cheongju, 28644, Rep. of Korea
[4]e-mail: lch5717@kaist.ac.kr
[5]e-mail: minkyo_seo@kaist.ac.kr

**Abstract**

Abstract: Allowing a high quality factor (Q-factor) to a sub-wavelength dielectric resonator, quasi-bound states in the continuum (Q-BIC) has gained much interest. However, the Q-BIC resonance condition is too sensitive to the geometry of the resonator and its practical broadband generation on a single-wafer platform has been limited. Here, we present that employing the base angle as a structural degree of freedom, the truncated nano-cone resonator supports the Q-BIC resonance with a high Q-factor of >150 over a wide wavelength range of >100 nm. We expect our approach will boost the utilization of the Q-BIC resonance for various applications requiring broadband spectral tuning.

**Introduction**

The bound states in the continuum (BIC) originally address the electron states bounded in a potential continuum[1]. The concept of the BIC has been adopted in photonics employing various nano-/micro-scale structures[2–9]. Recently, the discovery of a unique optical resonance with a Lorentzian peak shape in a high-index sub-wavelength dielectric resonator has led to the creation of the term quasi-BIC (Q-BIC). Q-BIC resonance arises from the hybridization between two azimuthally polarized modes with different modal properties, reducing radiative scattering losses. In comparison to typical Mie resonators, Q-BIC resonators have the advantage of a large Purcell factor due to their high Q-factors and small footprints. Q-BIC resonances have been intensively utilized for nanophotonic applications such as efficient second/third harmonic generation, bio- and heat-sensing, and lasing [7,10–16].

However, supporting the Q-BIC resonance at a specific wavelength requires strict constraints on the selection of the height and cross-section of the employed nanostructure[4,17]. It is not practical to control the vertical and lateral parameters independently with the conventional top-down wafer fabrication processes. Overcoming the constraints is the key to utilize the full potential of Q-BIC resonances across a broad wavelength range and enable wavelength-sensitive applications such as resonant Raman/photoluminescence spectroscopy and high-precision bio-sensing [13,18–20]. Demonstrating the Q-BIC modes in the telecommunication wavelength range is also a significant challenge due to the constraints.

In this work, we theoretically propose a promising method for broadband generation of Q-BIC resonances on a single-wafer platform. Although fixed in its height, the silicon (Si) truncated nano-cone resonator enables us to generate the Q-BIC condition over the entire range of the short-band (S-band) telecommunication wavelengths by employing the base angle as an additional degree of freedom (DOF). The demonstrated Q-BIC resonance supports the symmetric Lorentzian reflectance spectrum with a high Q-factor of >150 and a peak depth of ~0.6.

Figures 1(a) and 1(b) illustrate conventional nano-disks and truncated nano-cones that generate Q-BIC resonances with a fixed height. The nano-disk/cone resonator supports two azimuthal modes, of which the coupling creates the Q-BIC resonance. The $SiO_2$/indium tin oxide (ITO)/$SiO_2$ substrate boosts the coupling of the two azimuthal modes necessary to support the Q-BIC resonance [3,4]. The Breit-Wigner-Fano formula with respect to the asymmetric factor ($q$) characterizes the frequency-dependent reflectance (R) of the nano-resonator:

$$R(\omega) = A_0 + A_{\text{peak}} \frac{\left(q\, \Delta\omega + (\omega - \omega_c)\right)^2}{q^2(\Delta\omega^2 + (\omega - \omega_c)^2)}$$

Here, $\omega_c$, $A_{\text{peak}}$, $A_0$, and $\Delta\omega$ are the center frequency, peak depth, baseline, and linewidth of the Breit-Wigner-Fano line shape, respectively (Fig. 1(c)). The Q-BIC resonance with a Lorentzian shape corresponds to the case when q diverges to infinity.

The reflectance spectrum clarifies the distinct performance between the nano-disk and truncated nano-cone resonators. We calculated the reflectance spectra of 580-nm-tall Si nano-disk and truncated nano-cone resonators with various diameters, using the finite-difference time-domain (FDTD) method (Figs. 1(d) and 1(e)). The azimuthally polarized beam was injected with a numerical aperture (NA) of 0.6. The reflectance is given by the power ratio of the reflected light within the NA to the incident light. In Fig. 1(d), the diameter change from 800 nm to 880 nm causes the reflectance spectrum lineshape to gradually transition from a positive Fano resonance ($q < 0$), a symmetric Q-BIC resonance ($|q| \to \infty$) to a negative Fano resonance ($q > 0$). The Si nano-disk with a fixed height supports the Q-BIC resonance at a specific wavelength, with all other wavelengths exhibiting Fano resonances, regardless of diameter changes. On the other hand, the Si truncated nano-cone with an additional DOF (base angle) generates the Q-BIC resonance over a broadband wavelength range from 1466 nm to 1531 nm (Fig. 1(e)). It is worth noting that in practical fabrication processes, controlling the vertical shape of the etching mask, thermal treatment, and etching selectivity enables to efficiently engineer the base angle and lateral profile practically [21–27].

To understand the broadband generation of Q-BIC resonances in the truncated Si cone resonators, we analyzed the modal properties depending on the base angle and base diameter using the finite element method (FEM) simulation (Figure 2). The nano-disk/cone resonator supports two azimuthal modes, the axial and radial modes, which show an anti-crossing behavior as the base diameter varies (Fig. 2(a)). In this case, the azimuthal mode at the longer wavelength has a Q-factor lower than the other at the shorter wavelength. Hybridization of the axial and radial modes leads to the Q-BIC condition, where the hybrid mode at the lower wavelength has a significantly increased Q-factor and symmetric Lorentz line shape. In the truncated nano-cone, the resonance wavelength of the radial mode changes significantly depending on the base angle, while the resonance wavelength of the axial mode is determined primarily by the base diameter. Thus, the combined control of the base diameter and angle enables us to tune the wavelength of the Q-BIC condition widely over a broad spectral range.

Figures 2(b) and 2(c) present the Q-factor of the short- and long-wavelength azimuthal modes as a function of the base angle and base diameter of the truncated nano-cone, respectively. The dashed lines indicate the peak (Fig. 2(b)) and minimum (Fig. 2(c)) of the Q-factor distribution in the parametric space of the base angle and base diameter. The peak and minimum in the parametric space coincide, which represents that the mode hybridization makes the Q-factor of the short-wavelength (long-wavelength) mode further higher (lower). The emergence of mode hybridization at smaller diameters as the base angle decreases leads to the Q-BIC condition at a shorter resonance wavelength. In addition, Figure 2(d) shows the electric field intensity profiles of the azimuthal modes at a representative base angle of 74°. It is clearly observed that as the diameter changes, the mode profile evolves from the axial, hybrid, to radial mode and vice versa.

To confirm the Q-BIC resonance generation in the truncated Si nano-cone rigorously, we examined the asymmetric factor ($q$) in the two-dimensional map as a function of the base diameter and base angle (Fig. 3(a)). Fitting the reflectance spectrum calculated by the FDTD simulation with the Breit-Wigner-Fano formula yielded the asymmetric factor. The ideal Q-BIC condition is determined when the asymmetric factor diverges. The smaller the base angle, the smaller the diameter of the truncated nano-cone resonator to support the Q-BIC resonance. Even though they show similar behavior depending on the base angle and diameter, the conditions for maximizing the Q-factor of the hybrid mode and generating the Q-BIC resonance with a symmetric Lorentzian spectrum line are distinct. For a given base angle, the Q-BIC condition ($|q| \to \infty$) emerges at a smaller diameter than the condition that maximizes the Q-factor (white dashed line).

In Figure 3(b), the reflectance spectrum of the Q-BIC condition is plotted for different base angles from 66° to 90°. The Q-BIC resonance of the truncated nano-cone spans the entire range of the S-band telecommunication wavelengths, supporting the Lorentzian line shape. Our approach can be easily adapted to cover other optical wavelength bands such as O-, E-, C-, and L-bands, by adjusting the material and distance of the bottom mirror and tuning the height of the resonator [4]. The narrow spectral linewidth and symmetric line shape are beneficial to make the Q-BIC resonator applicable to various sensors in infrared wavelengths [18,19,28–30]. In addition, the large peak depth and high baseline allow a high dynamic range, making the Q-BIC resonator a promising element for use in efficient optical detectors [31].

Figure 3(c) summarizes the characteristic parameters of the Q-BIC resonance depending on the base angle of the truncated nano-cone. While the resonance wavelength is tuned from 1526 nm to 1423 nm, the Q-factor changes from 243 to 161. Over a wide wavelength range of >100 nm, we achieved a Q-factor of >150, sufficiently high for applications such as stimulated Raman scattering, low-threshold lasing, and optical nonlinear effect [10,14,15,20,32]. The Q-factor can be further improved by optimizing the structure and material of the bottom mirror [10,33]. Meanwhile, the peak depth remains at a large value of ~0.6, which indicates that the extinction cross-section and resonance strength of the truncated nano-cone resonator are robust regardless of the change in the base angle. These results clarify that the base angle is a practical DOF to for customizing Q-BIC resonances in sub-wavelength dielectric resonators.

In summary, we present a promising way to generate high-performance Q-BIC resonances over a broad wavelength range on a single-wafer platform. Employing the base angle as a structural DOF instead of the height, the Ssi truncated nano-cone resonator effectively controls the dispersion properties and hybridization of the azimuthal modes participating in the Q-BIC condition. We theoretically demonstrated the tuning of the symmetric Lorentzian reflectance spectrum with a high Q-factor and large peak depth across the S-band telecommunication wavelength range. The practical broadband generation of the Q-BIC resonance can be easily extended to other high-index functional materials, including III-V semiconductors and lithium niobite [34–36]. We thus expect that tailoring the wavelength of the Q-BIC resonance will be a promising technique for demonstrating various photonic devices such as lasers, modulators, and photodetectors with sub-wavelength-scale footprints.

**Figure legends**

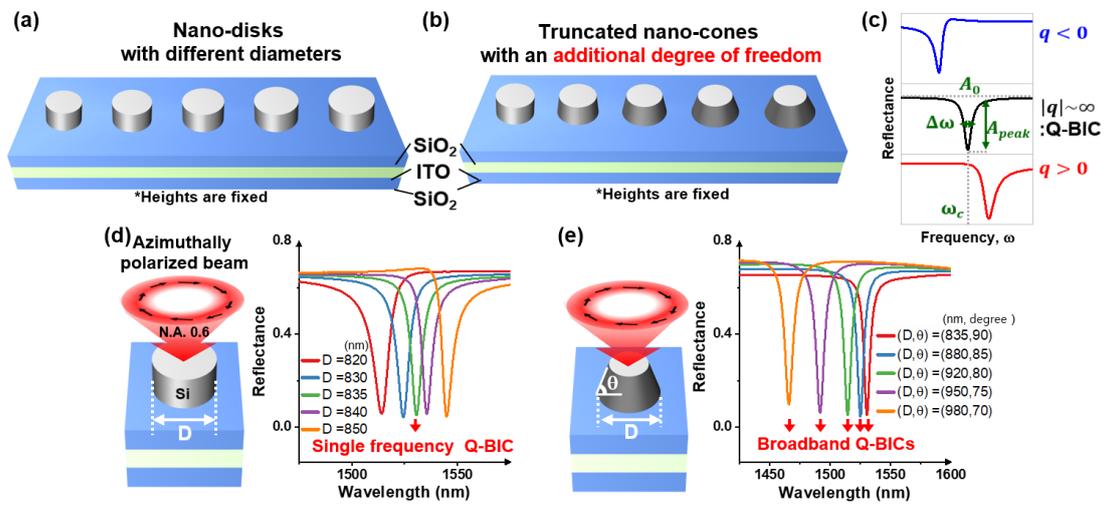

**Fig 1.** Generation of Q-BIC resonance in a subwavelength Si resonator. (a) Si nano-disks with one structural DOF (diameter). (b) Si truncated nano-cones with two structural DOFs (base angle and base diameter). (c) Schematic of the reflectance spectra of the Breit-Wigner-Fano line shape depending on the asymmetric factor ($q$). (d) Schematic diagram of the FDTD simulation and calculated reflectance spectra of Si nano-disk resonators with different diameters. The Q-BIC resonance is achieved only at a diameter of 835 nm. (e) Schematic diagram and calculated reflectance spectrum of the truncated nano-cone resonator. Control of the base diameters (D) and base angles (θ) enables the truncated nano-cone resonator to support Q-BIC resonances over a broadband wavelength range.

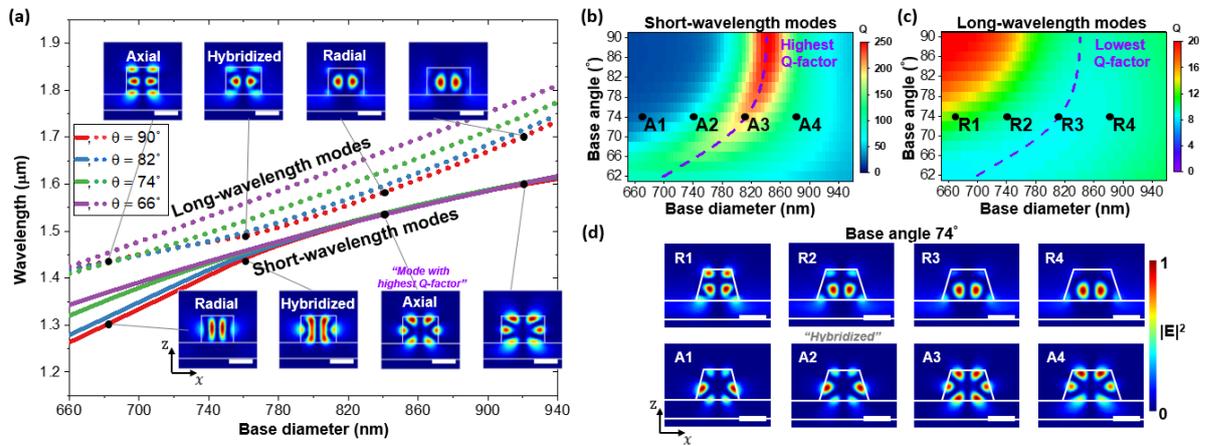

**Fig 2.** Hybridization of two azimuthal modes and emergence of the Q-BIC condition. (a) Resonant wavelength of the azimuthal modes depending on the diameter and base angle of the truncated nano-cone. (Inset) the electric field intensity profiles of the azimuthal modes in the nano-disk resonator (θ = 90). Two-dimensional map of the Q-factor of the axial (b) and radial (c) mode as a function of the diameter and base angle. The purple dashed line indicates the angle and diameter to support the condition maximizing(minimizing) the Q-factor of the short-(long-)wavelength mode. (d) The electric field intensity profiles of the axial (A1~A4) and radial (R1~R4) modes of four representative nano-cone resonators, of which the base diameter and base angle are indicated in (b) and (c). Scale bar, 500 nm

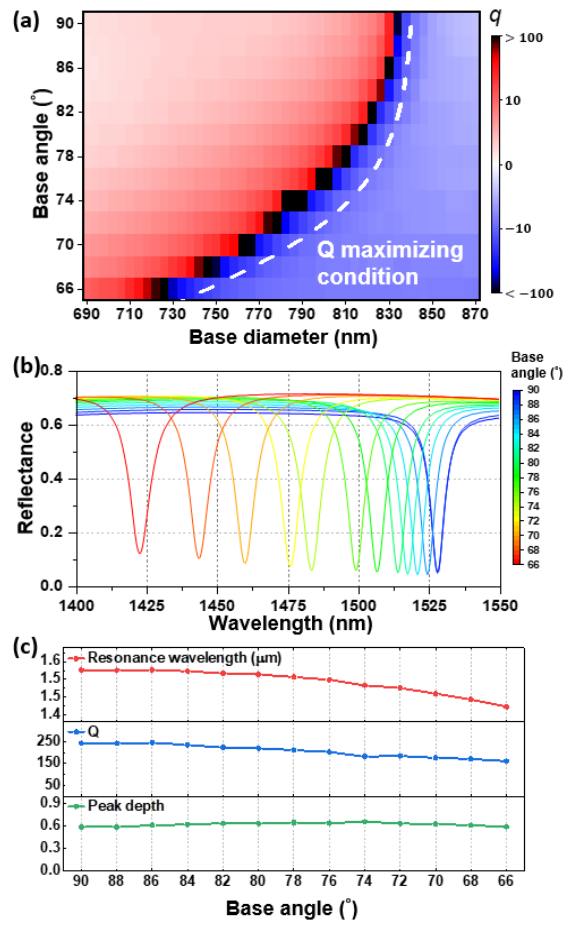

**Fig 3.** (a) Two-dimensional map of the asymmetric factor (*q*) as a function of the base angle and diameter of the truncated nano-cone resonator. The white dashed line indicates the condition maximizing/minimizing the Q-factor (See Fig. 2(b) and 2(c)). (b) Broadband generation of the Q-BIC resonance. (c) Calculated resonance wavelength, Q factor, and peak depth of the reflectance spectrum at the Q-BIC condition depending on the base angle. The azimuthally polarized beam injected with a NA of 0.6 was employed.